# STATISTICAL ANALYSIS FOR PRECISE ESTIMATION OF STRUCTURAL PROPERTIES OF NGC 1960

**Gireesh C. Joshi[1,2], R. K. Tyagi[2]**

**Abstract:** The statistical analysis of gathered of astronomical objects (such as open cluster) are excellent tools to compute its parameters and to constrain the theory of its evolution. Here, we are represented detailed structural and membership analysis of an open cluster NGC 1960 through various statistical formulas and approaches. The King empirical method provides the information about the cluster extent as 5.2 ± 0.4 pc. The exact identification of members of cluster is needed to precise estimation of its age, reddening, metallicity etc., therefore, to identify most probable members (MPMs), we adopted the combined approach of various statistical methods (kinematic, photometric and statistical) and 712 members are satisfied needed criteria of MPMs. The basic physical parameters of the cluster such as E(B−V)=0.23±0.02 mag, E(V−K)=1.05±0.03mag, log(Age)=7.35±0.05, and (m−M)=11.35±0.10 mag are obtained using the color-color and color-magnitude diagrams. NGC 1960 is found to be located at a distance of 1.34 ± 0.03 kpc. Using the archival proper motion catalogues, we estimate mean proper motions of NGC 1960 as 0.83±0.26 mas $yr^{-1}$ and -6.55±0.23 mas $yr^{-1}$ in the direction of RA and DEC, respectively.

**Introduction:** Open star clusters (OSCs) are studied to understand the evolution of galactic arms. A large fraction of stars in the direction of clusters does not belong to the family of cluster therefore knowing the membership of a star is an important means to precisely calculate its basic physical parameters with colour-magnitude diagrams (CMDs) and color-colour diagrams (CCDs). In the present paper, we are representing the structural parameters of NGC 1960. Its previous studies are summarized as follows. This cluster is situated in the Constellation Auriga and has been studied in the past by many authors. The spatial parameters with the dynamical behavioir of this cluster has been prescribed in the form of various previous work such as proper motion study [1]-[2], photoelectric study [3], photographic study [4] and CCD photometric study [2,5,6]. It is an intermediate age OSC and most massive (brighter) stars are found in its core region. The center of NGC 1960 is given as $\alpha_{2000}$ =05:36:18, $\delta_{2000}$ =+34:08:24 in the WEBDA archive. After these studies, it is needed to study of individual stars for improving the knowledge of the dynamics and evolution of this system. The characteristic of individual star would be well examined through time series observations. Such study leads to search variable stars within the OSC. This paper is organized as follows: The observational and reduction techniques of the standardization process of standard night are prescribed in Sec. 2. The center and radius of cluster are determined Sec. 3. The mean proper motion of the cluster is described in Sec. 4. The detailed membership analysis and most probable members are discussed in Sec. 5. The precise structural results of the cluster are shown in Sec. 6. The discussion and conclusions are described in Sec. 7.

**Detail of Standard Night:** The photometric observations of the OSC NGC 1960 including bias, twillight frame and two Landolt's standard field (SA95 and PG0231+051, [7]) were obtained through Johnson-Cousin UBVRI passbands using 1.04-m telescope at Manora Peak, Nainital on the night of 30 November 2010. We used 2k × 2k CCD (2×2 binning mode for improving the signal-to-noise ratio) camera covering a field of view of ~13×13 $arcmin^2$ of the sky. The read-out noise and gain of the CCD per pixel are 5.3 $e^-$ and 10 $e^-$ $ADU^{-1}$, respectively. A total of ten object frames was obtained having two frames each in U, B, V, R and I filter with exposure times of 300, 300, 200, 200 and 60-sec, respectively. The basic steps of image processing (bias subtraction, flat fielding and cosmic-ray removal) and photoelectric anlyasis were performed through the IRAF and DAOPHOT-II profile fitting technique [8], respectively. The instrumental magnitude was translated into standard magnitude using the transformation equation:

$$m_i = M_i + z_i + c_i \times colour + k_i \times X \qquad (1)$$

where $z_i$, $c_i$, $m_i$, $M_i$, $k_i$ and X represent the zero-point, colour-coefficient, aperture instrumental magnitude, extinction coefficient and air-mass, respectively. The (U−B), (B−V), (V−R) and (R−I) colours were used to determine instrumental magnitudes in U, B, V, R and I passbands respectively (details available in Table 01 and standard residual with photometric error has been depicted in Fig. 01.). We have found calibrated standard deviation of standard stars as 0.083, 0.071, 0.047, 0.030 and 0.049 mag in U, B, V, R and I filters, respectively.

**Table 01**: *The zeropoint, colour-coefficient and extinction-coefficient of different passbands.*

| Filter | Zeropoint($z_i$) | Colour coefficient($c_i$) | Extinction Coefficient($k_i$) |
|---|---|---|---|
| $U$ | 8.16 ± 0.01 | -0.05 ± 0.01 | 0.55 ± 0.02 |
| $B$ | 5.81 ± 0.02 | -0.01 ± 0.02 | 0.29 ± 0.03 |
| $V$ | 5.43 ± 0.01 | -0.08 ± 0.01 | 0.15 ± 0.01 |
| $R$ | 5.23 ± 0.01 | -0.09 ± 0.02 | 0.09 ± 0.02 |
| $I$ | 5.63 ± 0.02 | 0.01 ± 0.01 | 0.07 ± 0.02 |

The pixel coordinates (x,y) of CCD images are transformed into the celestial coordinates ($\alpha_{2000}$, $\delta_{2000}$) by utilizing *CCMAP* and *CCTRAN* tasks of *IRAF* (for this, a set of 63 stars is obtained through skycat software). We compared our photometry with wide field (~50′×50′) UBVRI photometry of NGC 1960 ([5], they were using the 105-cm Kiso Schmidt telescope). This comparision gives 1038 common stars (within a rms scatter of 1 arc-second) and further reveal that our photometry matches well with UBVR bands and shows a linear trend and shift between them in the I band (as depicted in Fig. 2). The IR photometric magnitudes of stars were extracted from the two-micron all sky survey (2MASS) [9] and Wide-field mid-IR Survey Explorer (WISE) [10] surveys within 1 arcsec rms deviation of stellar positions estimated by UBVRI photometry. Our photometry resulted in a total of 1605 stars within 13′×13′ field of the cluster NGC 1960 in which 447, 1088, 1424, 1583 and 1532 stars were found in U, B, V, R and I bands, respectively. We have combined UBVR magnitudes of both catalogue and took previous I-magnitude due to poor quality of new I band images.

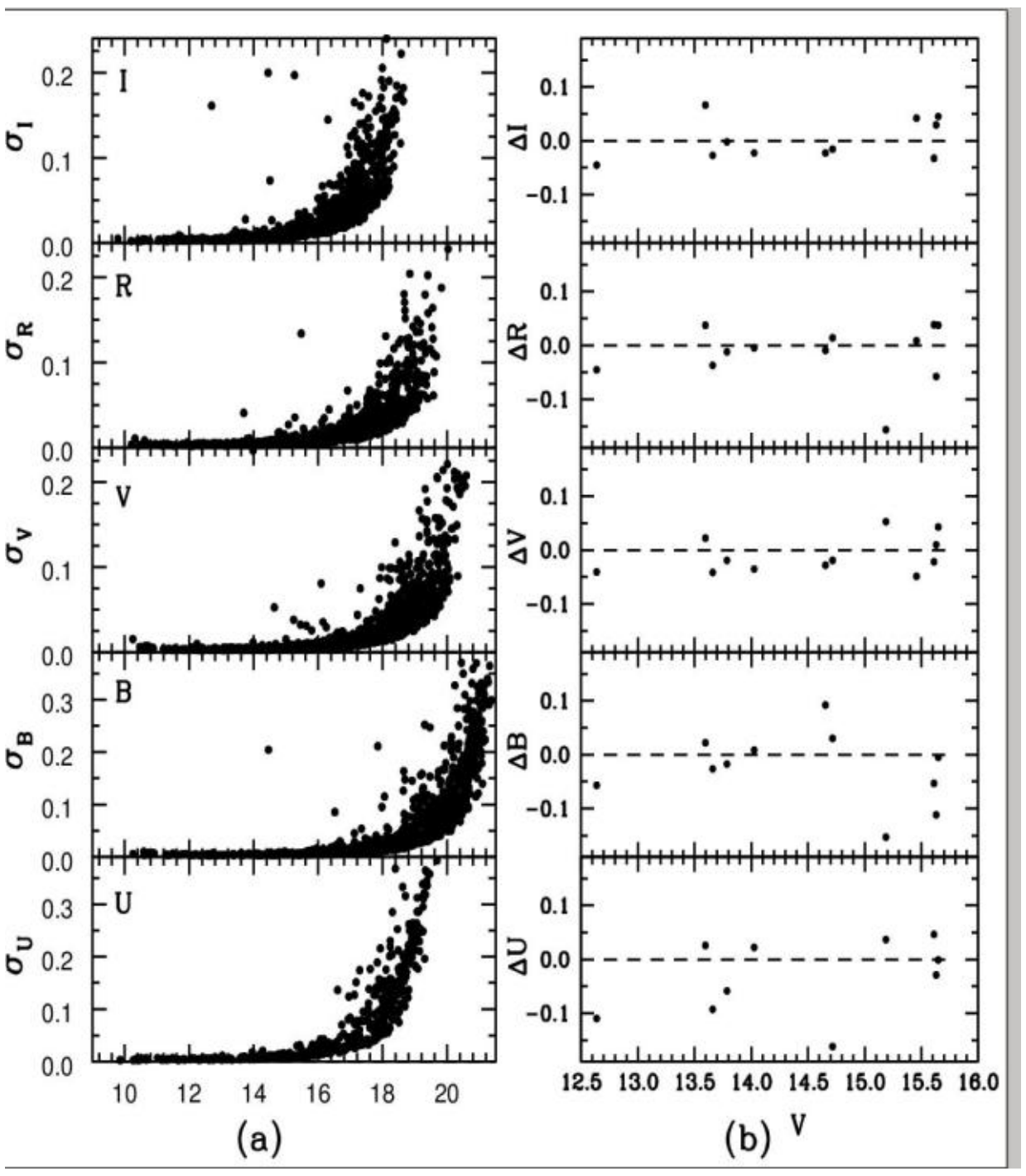

**Figure 01**: ***a)***- the standard deviation (errors) of stars as a function of *brightness.* ***b)****- difference between our estimated magnitudes with that of the Landolt's magnitude for the standard stars in UBVRI passbands. The black dashed line represents zero shifts.*

**Center and Radius of the cluster:** There are many extremely bright stars close to the cluster center which are either saturated or contaminated by the nearby stars in our observed frames. This resulted in under-estimation of the stellar density in immediate neighborhood of the cluster center. Hence, we only used the cluster center given in the WEBDA. The radial density profile (RDP) of open cluster is used to determine the cluster radius. The RDP is constructed by determining the stellar density (for stars brighter than V=19 mag) in concentric rings of equal width (here, 1arcmin or ~ 80 pixels) around the cluster center.

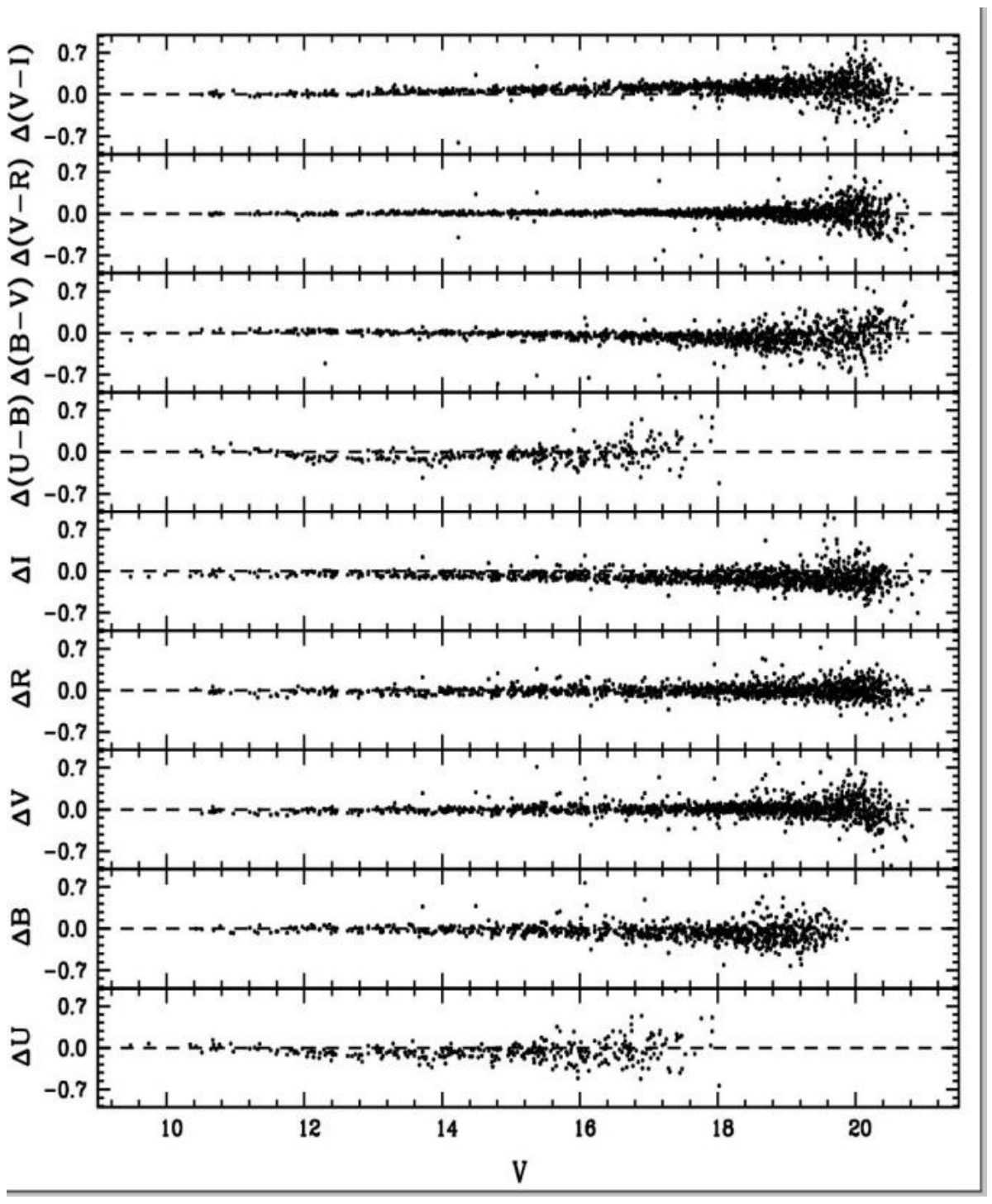

**Figure 02**: Comparison of the present photometry to the photometric study by Sharma *et al. [13]. The horizontal dashed line in each panel represents the zero magnitude difference between two catalogues.*

The stellar density ($\rho_r$) at the distance r from the cluster center is defined by the following empirical formula [11]:

$$\rho(r) = \rho_f + \frac{\rho_0}{1+\left(\frac{r}{r_c}\right)^2} \qquad (2)$$

where $\rho_0$ is the peak stellar density and $\rho_f$ is the background density. The $r_c$ is the core radius of the cluster defined as a distance from the center at which the density ρ(r) becomes half of the $\rho_0$. A sharp drop in the stellar density within 1 arcmin radius around the cluster center is seems to be due to some stars near the cluster center that are very close to few other extremely bright star and may be unresolved in the present photometry. We therefore excluded the very first point in the RDP and a best fit (continuous line in Fig. 4-a) was obtained from the remaining points through $\chi 2$ minimization technique. We estimated the core and cluster radii (RDP intercepts to background density) of the cluster NGC 1960 as 4.2±0.9 and 16±1 arcmin, respectively. A similar radial density distribution has also been drawn for 2MASS J and K bands for the same set of stars as used in V band but for which J and K magnitudes are available in the 2MASS catalogue (depicted in figs. 4-b & 4-c).

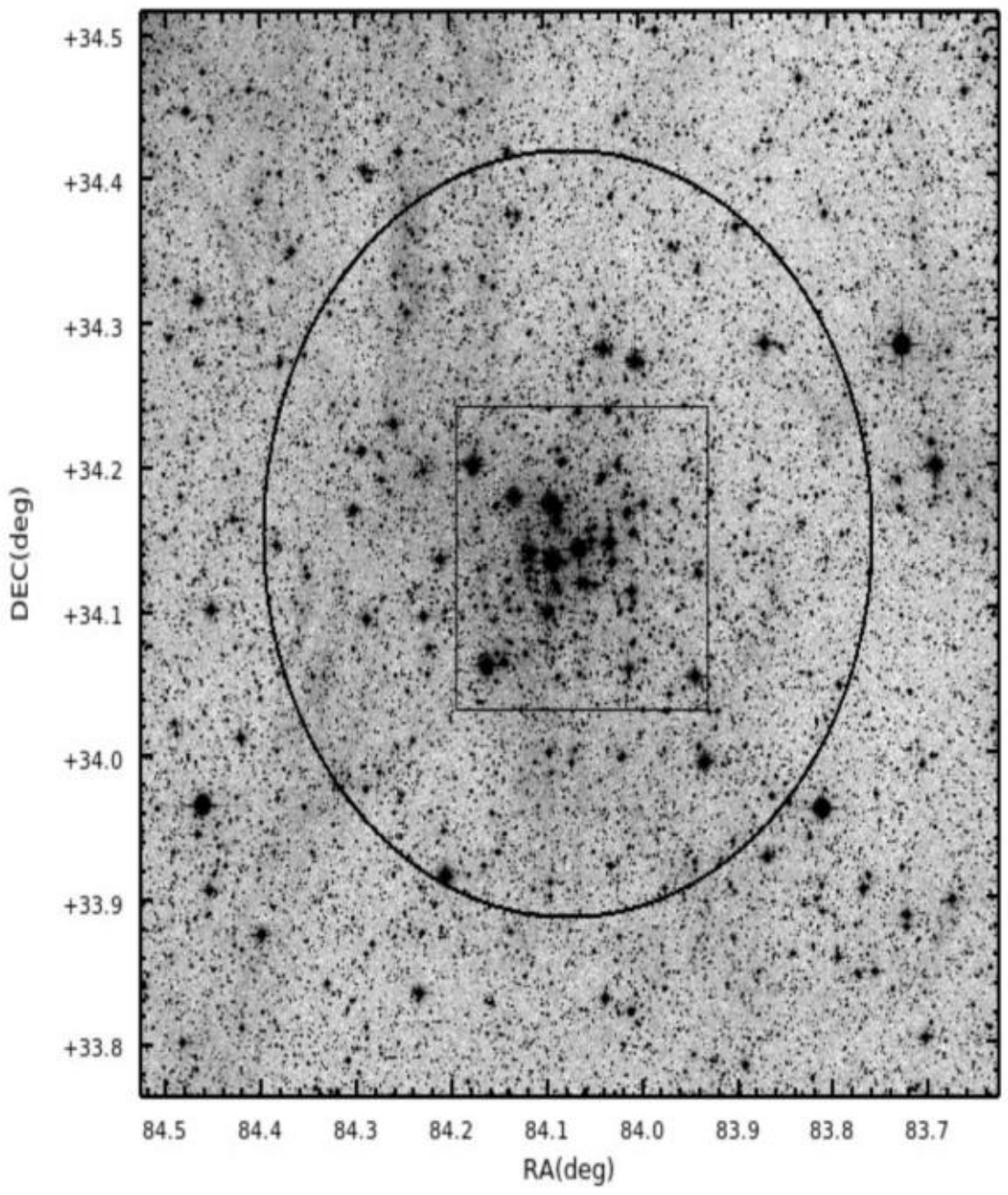

***Figure 03****: The finding chart of stars in the cluster field of NGC 1960. The circle represents the cluster extent and square region shows the field of view of our telescope.*

For a constant background density, we also found a slightly smaller cluster radius in comparison to the V-band's radius. However, it is not unusual to have slightly different radius in different wavebands or cut-off magnitudes as cluster radius determined through RDP depends on the completeness of data and visibility of the fainter stars in the observed frames. The cluster extent as determined in the present study is marked by circle on the DSS frame as illustrated in Fig. 3.

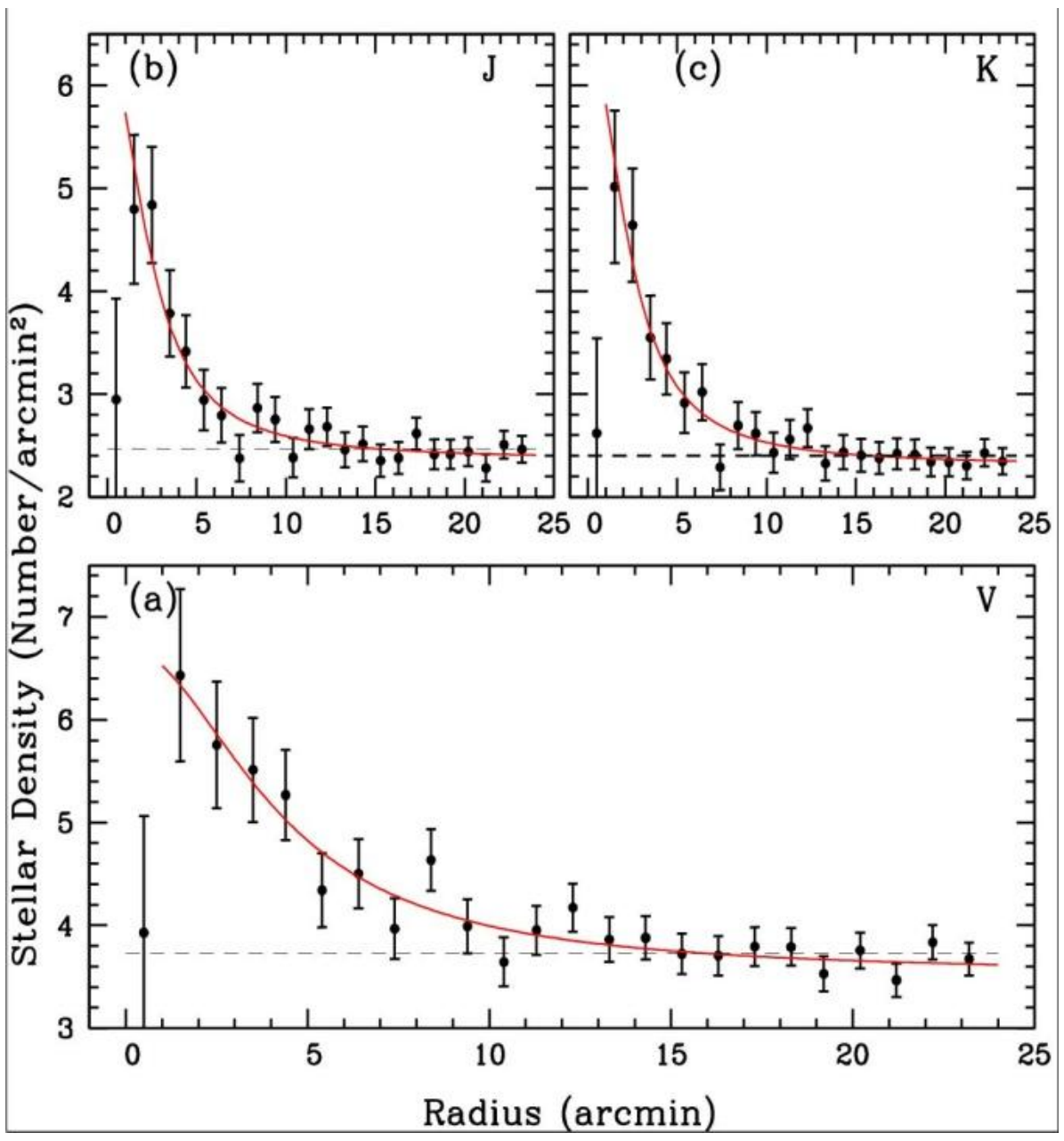

*Figure 04*: *The radial density profile for NGC1960 in (a) V band, (b) J band, and (c) K band. The solid line represents the King profile while the horizontal dashed line indicates the field density.*

**Mean proper motion of cluster:** For determination of the mean proper motion of the cluster, we used UCAC4 proper motion catalogue [12] and PPMXL catalogue [13]. PPMXL is a catalogue of positions, proper motions, 2MASS and optical photometry of 900 million stars and galaxies down to V~21 mag, while UCAC4 catalogue is available mostly for relatively brighter stars. A comparison of both catalogues shows large discrepancy in proper motion values and UCAC4 catalogue show better accuracy of proper motion values than the PPMXL catalogue. Hence, we combined both the catalogues to give preference to the UCAC4 catalogue of common stars in both catalogue and selected stars should not have error more than 5 mas/yr in the proper motions. After comparing our catalogue with the combined proper motion catalogue, we found 9048 common stars, in-which 2964 stars fall within the cluster radius. The distribution of these stars in $\mu_x$-$\mu_y$ plane is shown in Fig. 5. For more accurate proper motion, we adopted V=14mag as cut-off to determine of mean proper motion of the cluster and this cut-off provides 238 stars within the cluster. The mean and standard deviation of the proper motions of these stars were computed and stars with proper motions outside ± 2σ were rejected. This procedure was repeated until no more stars were rejected. The remaining 111 stars (shown by dark filled circle in Fig. 5) gives mean proper motion as:- $\mu_x = -0.08 \pm 0.11$ mas/yr, $\mu_y = -5.41 \pm 0.11$ mas/yr.

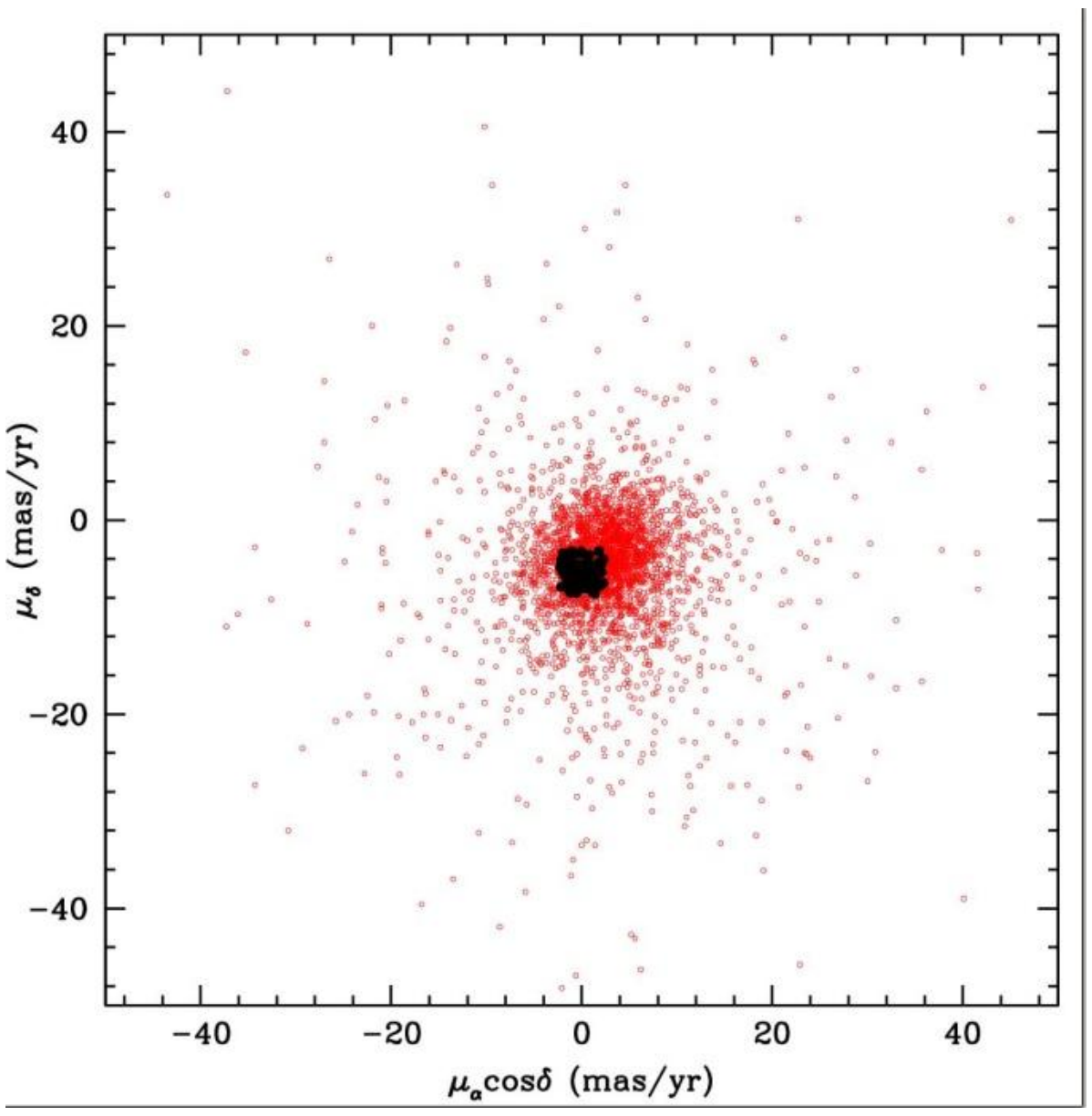

*Figure 05*: *The distribution of proper motion of stars. The large points represent those stars which are used to determine the mean proper motion of the cluster.*

**Membership probabilities and most probable members:** A large fraction of detecting stars does not belong to the cluster. Since members of the open clusters have the same age, distance, proper motion and spatial velocity, hence they follow a pattern in the colour-magnitude diagrams and proper motion plane. In order to identify the members of cluster, we have derived three different probabilities based on spatial, kinematic and photometric which are explained below.

- ***Spatial probability***: The spatial probability, $p_{sp}$, is related to the angular position of the stars with respect to the cluster center and is derived by following relation:

$$p_{sp} = 1 - \frac{r}{r_c} \qquad (3)$$

where r is the angular distance of stars from the cluster center and $r_c$ is the cluster radius and $p_{sp} > 0$ for stars $r<r_c$.

- ***Kinematic probability:*** The deviation in the proper motion of stars in both RA and DEC directions with respect to the mean proper

| $ID$ | $V$ (mag) | $e_V$ (mag) | $B-V$ (mag) | $e_{B-V}$ (mag) | $p_{sp}$ | $p_{ph}$ | $p_k$ |
|---|---|---|---|---|---|---|---|
| 16 | 9.485 | 0.022 | 0.034 | 0.040 | 0.87 | 1.00 | 0.80 |
| 18 | 9.600 | - | 0.050 | - | 0.46 | 1.00 | 0.91 |
| 20 | 9.740 | 0.024 | 0.021 | 0.047 | 0.88 | 0.98 | 0.67 |
| 34 | 10.390 | 0.060 | 0.010 | 0.078 | 0.58 | 0.85 | 0.75 |
| 42 | 10.428 | 0.007 | 0.101 | 0.028 | 0.86 | 1.00 | 0.55 |
| . | . | . | . | . | . | . | . |
| . | . | . | . | . | . | . | . |

motion of the cluster is known as kinematic probability. The kinetic probability, $p_k$, has been derived by following relation [14]:

$$p_k = exp\left\{-0.25\left[(\mu_x - \bar{\mu}_x)^2/\sigma_x^2 + (\mu_y - \bar{\mu}_y)^2/\sigma_y^2\right]\right\} \quad (4)$$

- ***Photometric probability:*** Photometric probability, $p_{ph}$, was derived with the reference to the blue and red sequence limits in (B−V)/V colour-magnitude diagram (CMD). For further analysis, we considered only those cluster stars which have proper motion within 1σ of the mean proper motion of the cluster. We fitted isochrones of solar mettalicity [15] in the (B−V)/V CMD through an iterative approach and best visual fit provides blue and red sequences of the cluster. The shift in magnitude and colour for red sequence were taken as -0.75 mag and 0.042 mag, respectively due to the unresolved MS binaries. We assigned photometric probability, $p_{ph}$, as 1 for 755 stars which laying between above said both sequences and said probability for remaining stars by given relation as:

$$p_{ph} = exp\left[-0.5 \times \left(\Delta(B-V)/\sigma_{(B-V)}\right)^2\right] \quad (5)$$

where Δ(B−V) is the colour differences, and σ(B−V) is the corresponding photometric errors in color.

- ***Most probable cluster members:***

The precise membership details of the stars in the cluster field are utterly important to determine the precise value of basic parameters in the cluster. Here, we assigned most probable members (MPMs) to those 712 stars which satisfy following criterion:

***1-*** *They are stars of the cluster i.e.* $p_{sp} > 0$ *and* $p_{ph} > 0.7$.

***2-*** *The proper motion of stars should lie within 1σ of the mean proper motion (*$P_k$*>0.5) of the cluster.*

We also note that some very bright stars are excluded from this list due to cause of either saturated or close enough to other bright objects. Identified MPMs is conservative due to their confer membership status, but we cannot deny membership of other stars. A sample of the MPMs catalouge is given in Table 2.

**Table 02**: *The sample list of photometric magnitude and colours of the 712 identified MPMs.*

**Basic parameters of NGC 1960:** We took advantage of our identified MPMs for re-deriving the cluster parameters.

- ***Reddening in optical bands:***

The (U−B) vs (B−V) TCD is used to determine the reddening i.e., E(B−V) through spectral type stars earlier than A0 due to fact that later type stars may be more effected by the metallicity and background contamination [16]. We draw zero age-main sequence i.e., ZAMS [17], for observed MS stars on the above said TCD (Fig. 06). We assumed a normal reddening vector i.e., E(U−B)/E(B−V)=0.72, and adopted a solar metallicity of cluster due to lack of information of metal content. The reddening error is calculated by using the following relation [18]:

$$\sigma^2_{E(B-V)} = \sigma^2_{(U-B)_0} + \sigma^2_{(B-V)_0} + \sigma^2_{vector} \quad (6)$$

where $\sigma_{vector}$=0.01 mag. The value of E(B-V) is obtained to be 0.23±0.02 mag by the visual best fit in the TCD, which is consistent with the 0.22 mag [13] and 0.25 ± 0.2 [2].

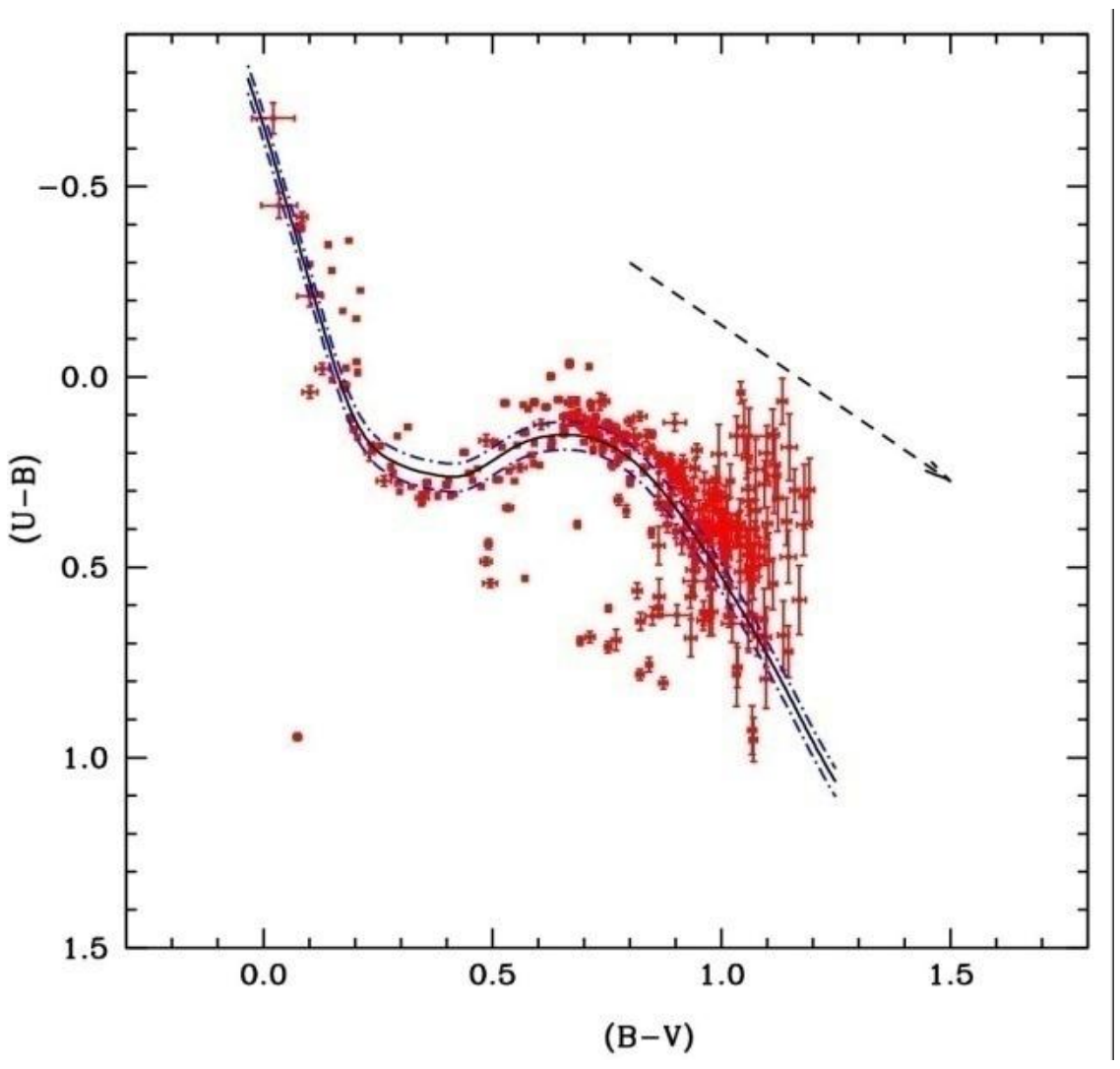

***Figure 06***. *(B−V) vs (U−B) 2-colour diagram.*

A closer look at the TCD indicates a non-negligible variable reddening in the cluster region with an error of ±0.12 in the slope of normal reddening vector E(U−B)/E(B−V) as shown by two dashed lines in Fig. 6. The (V-K) vs (J-K) diagram (depicted in Fig. 7) is used to determine the interstellar extinction in the near-IR range through overplotting Marigo isochrones of solar metallicity in it.

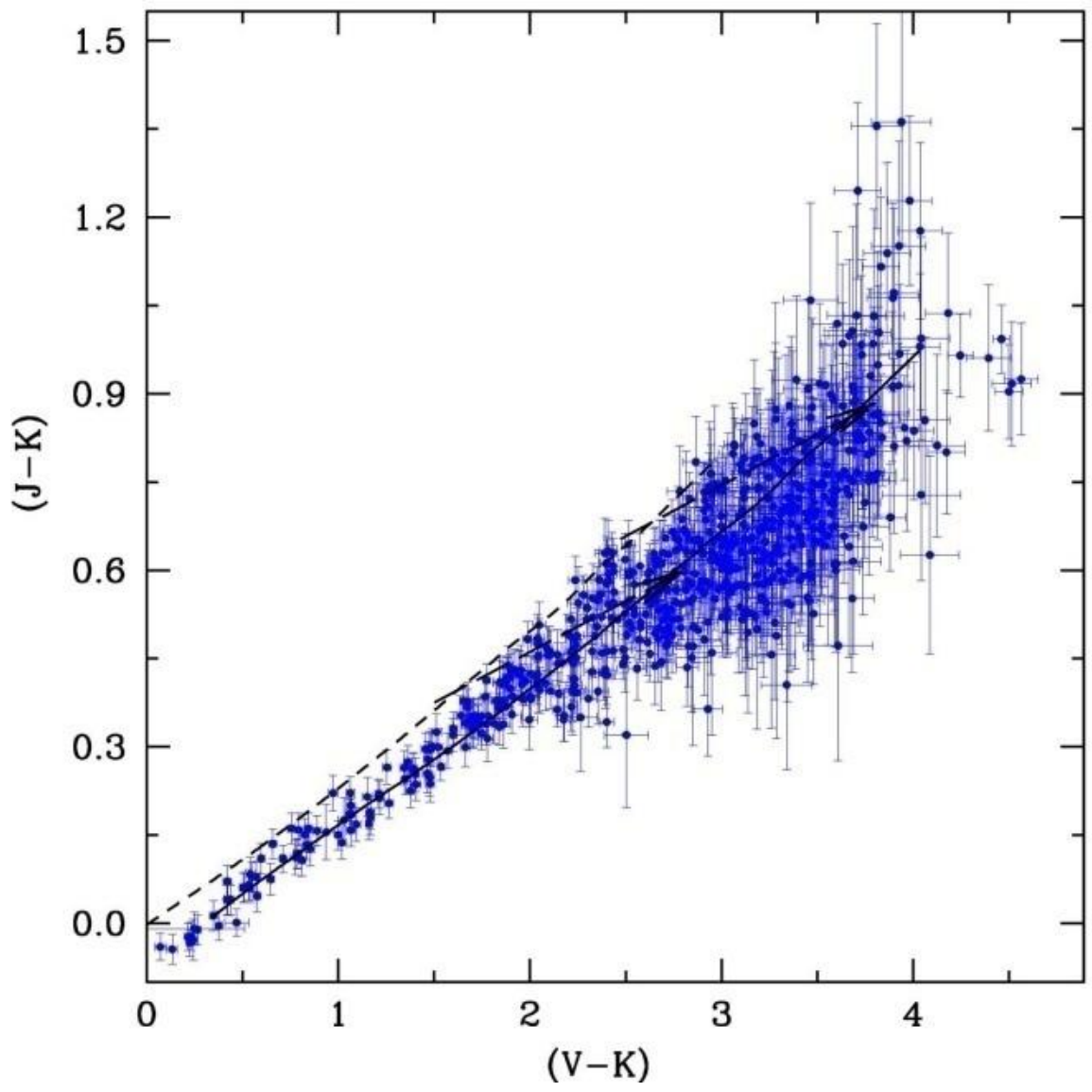

***Figure 07:*** *The (V-K) vs (J-K) diagram (dashed and continuous lines represent Marigo's isochrones as without shift and best fit respectively. Arrows represent the reddening vector E(J-K)/E(V-K)* = 0.173.

The best fitted isochrones provides the value of colour excesses of *E(V − K)* and *E(J − K)* as 1.05±0.03 mag and 0.18±0.01 mag respectively. These values provide the reddening value as 0.38±0.03 mag by following relation [19]:

$$R = 1.1 \times \frac{E(V-K}{E(B-V)} \qquad (7)$$

The new value of reddening indicates a large IR excess for stars in the cluster *NGC 1960*.

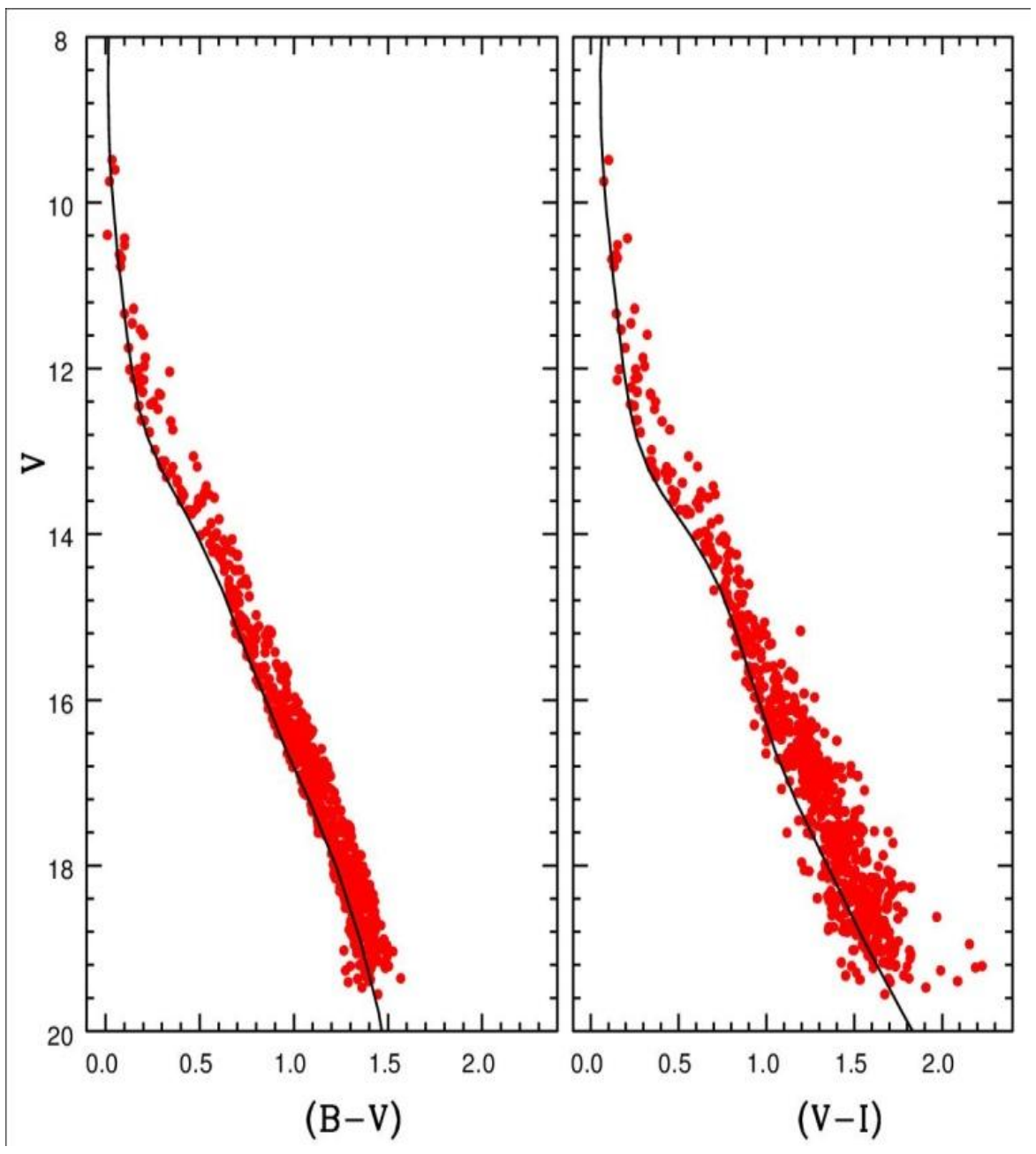

***Figure 08****: (B-V)/V and (V-I)/V CMDs in the field of cluster. The solid line represents the best fit isochrone to the cluster MS for log(Age)=7.35 and (m−M)=11.35 mag.*

- ***Age and Distance:*** The (B−V)/V and (V−I)/V CMDs for MPMs show a well populated, but broad main sequence (MS) stars that may be due to photometric errors and/or presence of binary stars within the cluster. Marigo's theoretical isochrones are over plot on the CMDs by varying the distance modulus and age simultaneously in both (B−V)/V and (V−I)/V CMDs while keeping reddening E(B−V)=0.23 mag and E(V−I)= 0.29 mag fixed (as depicted in Fig. 08). The best fit provides a distance modulus ($V-M_V$) =11.35±0.0.10 mag and log(Age) = 7.35±0.05 (yr). Furthermore, these values lead a true distance modulus $(m-M)_0$ and distance as 10.64±0.05 mag and 1.34±0.03kpc respectively (close to literature [5]).

**Conclusion:** Using a V-band data of the present catalogue for the stars brighter than V=19 mag, we found that the core and cluster radius are 4.2±0.9 and 16±1 arcmin, respectively. In a similar analysis of the 2MASS's J and K bands, we found a lower core and cluster extent for the cluster NGC 1960. The spatial, photometric and kinematic probabilities of the stars in the field of cluster are determined from their positions in the cluster field, located in the colour-magnitude diagram and proper motion, respectively. On the basis of these probabilities, a total of 712 MPMs is identified within the cluster. Using these MPMs, we derived the reddening in optical band using (U−B)/(B−V) diagram as 0.23±0.02 mag. Based on the (B−V)/V and (V−I)/V CMDs and a visual fitting of isochrones of solar metallicity to the blue sequence on the CMD, we determined a distance of 1.34±0.03 kpc and log(age) of 7.35±0.05. At this distance, we estimated a respective core and cluster radii of 1.6±0.3 pc and 5.2±0.4 pc for this cluster. The mean proper motion of the cluster was determined using the most recent proper motion catalogue (UCAC4 and PPMXL) and found to be −0.08±0.11 mas/yr and −5.41±0.11 mas/yr in the direction of RA and DEC, respectively. The left massive stars from MPMs list of present studied cluster indicated that it is further required to improve the procedure of identification MPMs.

**Acknowledgment**

GCJ is thankful to Mr. Anil Chauhan and Prabhat (AP cyberzone, Nanakmatta) for providing the computer facilities. GCJ is also acknowledged to ARIES, Nainital for providing observing facility duration Oct, 2012 to April 2015 and also acknowledge to Dr. Santosh Joshi with Dr. Brajesh Kumar for providing data of NGC 1960 on the night of 30Nov-2010.

* * *

Author 1 : P.P.S.V.M.I. College, Nanakmatta, U.S. Nagar, Uttarakhand-262311 (INDIA), /Researcher/ email:- gchandra.2012@rediffmail.com

*and*

Department of Physics, H.N.B. Govt. P.G. College, Khatima, U.S. Nagar, Uttarakhand- 262308 (INDIA) / Researcher

Author 2 : Department of Physics, H.N.B. Govt. P.G. College, Khatima, U.S. Nagar, Uttarakhand- 262308 (INDIA) / Associate Professor/ email: rajkumar.tyagi@gmail.com